\documentclass[aip,rsi,reprint,groupedaddress]{revtex4-1} 
\usepackage[utf8x]{inputenc}
\usepackage{graphicx}
\usepackage[per-mode=repeated-symbol]{siunitx}
\usepackage{xspace}
\usepackage{tikz}
\usetikzlibrary{shapes.geometric}

\usepackage{etoolbox}

\makeatletter
\appto\abstract{%
  \let\latexlist\list \rightskip=\leftskip
  \def\list{\edef\keeprightskip{\the\rightskip}\latexlist}%
  \patchcmd\latexlist{\ignorespaces}{\rightskip\keeprightskip\ignorespaces}{}{}%
}
\makeatother

\DeclareRobustCommand\redrectangle{\tikz{\node[rectangle,fill=red,scale=0.7] at (0,0) {};}}
\DeclareRobustCommand\greendiamond{\tikz{\node[diamond,fill=green!70!black,scale=0.5] at (0,0) {};}}

\newcommand{\Rb}[1]{\ensuremath{^{#1}\textrm{Rb}}\xspace}

\begin{document}
\title{A scalable high-performance magnetic shield for Very Long Baseline Atom Interferometry}

\author{E. Wodey}
\author{D. Tell}
\author{E. M. Rasel}
\author{D. Schlippert}\email{schlippert@iqo.uni-hannover.de}
\affiliation{Leibniz Universität Hannover, Institut für Quantenoptik, Welfengarten 1, 30167 Hannover, Germany}

\author{R. Baur}
\author{U. Kissling}
\author{B. Kölliker}
\author{M. Lorenz}
\author{M. Marrer}
\author{U. Schläpfer}
\author{M. Widmer}
\affiliation{IMEDCO AG, Industriestrasse West 14, 4614 Hägendorf, Switzerland}

\author{C. Ufrecht}
\affiliation{Universität Ulm, Institut für Quantenphysik and Center for Integrated Quantum Science and Technology (IQ$^\text{ST}$), Albert-Einstein-Allee 11, 89069 Ulm, Germany}

\author{S. Stuiber}
\author{P. Fierlinger}
\affiliation{Technische Universität München, Physikdepartment, 85748 Garching, Germany}

\date{\today}

\begin{abstract}
  We report on the design, construction, and characterization of a \SI{10}{\meter}-long high-performance magnetic shield for Very Long Baseline Atom Interferometry (VLBAI). We achieve residual fields below \SI{4}{\nano\tesla} and longitudinal inhomogeneities below \SI{2.5}{\nano\tesla\per\meter} over \SI{8}{\meter} along the longitudinal direction. Our modular design can be extended to longer baselines without compromising the shielding performance. Such a setup constrains biases associated with magnetic field gradients to the sub-\si{\pico\meter\per\second\squared} level in atomic matterwave accelerometry with rubidium atoms and paves the way towards tests of the universality of free fall with atomic test masses beyond the $10^{-13}$ level.
\end{abstract}

\maketitle

\section{Introduction}

Light pulse atom interferometers are powerful tools for modern precision metrology.\cite{Rosi2014,Bouchendira2011,Parker2018} They exploit the fine and well understood control of matter waves by light to achieve record instabilities and inaccuracies in the precision measurement of forces and other inertial quantities.\cite{Hardman2016,Barrett2016,Jaffe2017,Sabulsky2019,Freier2016,Gillot2014} In the conventional Kasevich-Chu geometry,\cite{Kasevich1991} the phase sensitivity of these interferometers scales linearly with the enclosed space-time area. By extending the free fall distance from tens of centimeters to the order of ten meters, large scale atom interferometers \cite{Kovachy2015,Zhou2011} target more than a factor fifty increase of their scale factor. When applied to the measurement of the local gravitational acceleration, the combination of such long baselines with high-performance inertial reference platforms brings short-term instabilities competing with state of the art superconducting gravimeters in reach, while in addition providing absolute measurements.

However, spurious field gradients along the free evolution path of the atoms can mimic the signal of interest and therefore limit the measurement accuracy and instrumental stability. In particular, magnetic field inhomogeneities generate bias accelerations on the atoms due to the Zeeman effect. Magnetic gradients at the few \si{\nano\tesla\per\meter} level or better along baselines of several meters are therefore required to satisfy the accuracy budget of these large scale atom interferometers at the sub-\si{\nano\meter\per\second\squared} level, compatible with their target instability.\cite{Hartwig2015} Magnetic shielding is typically achieved by channeling external magnetic flux inside a high permeability shell around the volume to isolate.\cite{Mager1970} Here, the homogeneity of the shielding material's permeability is key to ensure a uniform magnetization of the shield. This is however challenging on the lengths required by large scale atom interferometers since the production of homogeneous extended curved sheet metal is problematic, as well as the gap-free and reproducible junction of several pieces. A commonly used solution is to produce a fully welded assembly which is subsequently hydrogen annealed to ensure homogeneous properties of the shielding material.\cite{Dickerson2012} This is however impractical due to the limited availability of suitable furnaces and lacks scalability for future, possibly larger applications. Here, inspired by the layout of magnetically shielded rooms,\cite{Altarev2014} we report on the design, construction, and characterization of a \SI{10}{\meter}-long magnetic shield for the Hannover Very Long Baseline Atom Interferometry (VLBAI) facility.\cite{Hartwig2015} We first describe the key design points in achieving a fully length-scalable, large length-to-diameter ratio magnetic shield which does not require overall annealing. We then assess the shield's performance with residual field and dynamical shielding factor measurements and finally discuss its application in precision atom interferometry.

\section{Shield design and construction}

\begin{figure}
  \includegraphics[width=.45\textwidth] {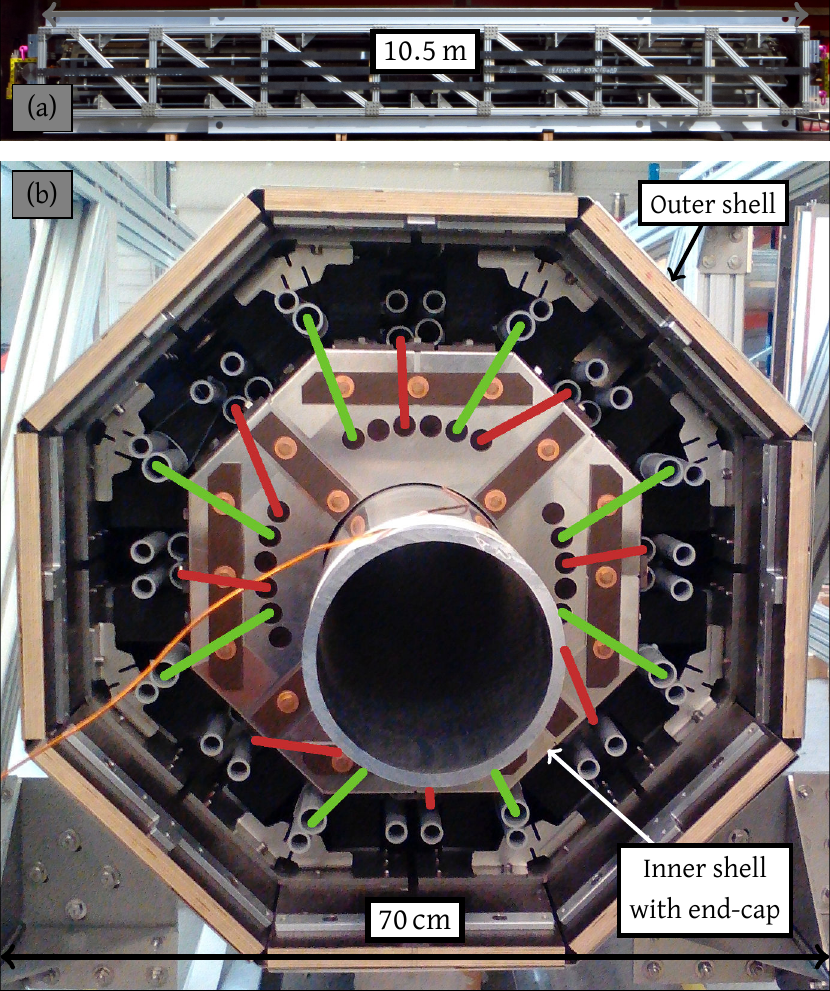}
  \caption{Photographs of the assembled magnetic shielding enclosure. (a) Side view of the final assembly. (b) End view with the end-cap for the outer shell removed. Two octagonal prism permalloy shells are supported by a plywood structure. Between the shells, PVC pipes guide the magnetic equilibration coils. Red and green lines show the routing of the two coils used for the equilibration of the inner layer. In the center, an aluminum pipe holds the space for the interferometer's vacuum chamber.}
  \label{fig:open-shield}
\end{figure}

In our atom interferometer, atoms fall freely inside a vertically oriented \SI{10.5}{\meter} long aluminum ultra-high vacuum pipe \footnote{Seamless extrusion of aluminum EN AW 6060, residual pressure below \SI{1e-9}{\milli\bar}} with inner diameter \SI{18}{\centi\meter} and outer diameter \SI{20}{\centi\meter}. The magnetic shielding enclosure for this cylindrical vacuum chamber consists of two concentric octagonal prism shells of high permeability Ni-Fe alloy \footnote{Krupp Magnifer 7904} (``permalloy'') as shown in figure~\ref{fig:open-shield}. The inner and outer permalloy shells have circumscribed diameters of \SI{450}{\milli\meter} and \SI{750}{\milli\meter} respectively. Both shells are closed individually by end-caps resulting in a total length of \SI{9.7}{\meter} for the inner shell and \SI{10}{\meter} for the outer one. The entire shielding assembly weighs around \SI{7500}{\kilo\gram}. On the main axis of symmetry, a \SI{22}{\centi\meter} diameter circular opening is left in the end-caps as clearance for the vacuum chamber.

\begin{figure}
  \includegraphics[width=.45\textwidth,trim=1 1 1 1,clip] {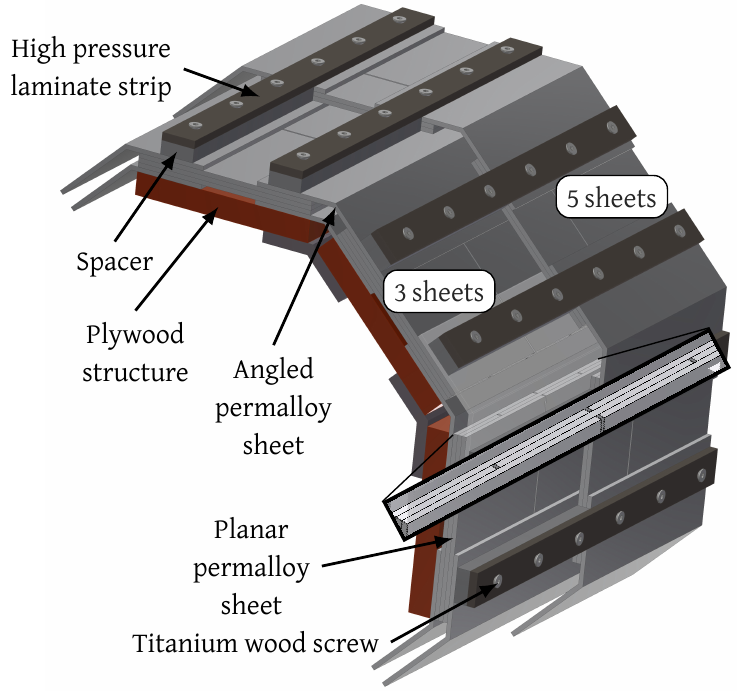}
  \caption{Principle for the assembly of the permalloy sheets for a single shell (not to scale). The \SI{1}{\milli\meter} thick permalloy sheets are stacked and pressed against a supporting plywood structure. Planar and angled strips are alternated. In the longitudinal direction, gaps between consecutive sheets (planar or angled) are closed by offseting the next sheet layer as shown in the inset. High pressure laminate (HPL) strips help distributing the load applied by the titanium wood screws. We use one screw every \SI{250}{\milli\meter} to ensure uniform pressure on the permalloy sheets. Since the total nominal thickness of the shield varies along its length, polymer spacers are used to keep the HPL strips level as shown here for the transition between three and five planar sheets (\SI{3}{\milli\meter} to \SI{5}{\milli\meter} effective shell thickness).}
  \label{fig:stackup-example}
\end{figure}

Despite the cylindrical symmetry of the volume to isolate, an octagonal geometry turned out to be more practical. It allows using mainly planar sheet material with well defined and homogeneous permeability after annealing which enables more reliable finite element simulation of the assembly. Figure~\ref{fig:stackup-example} shows the construction principle of the shielding enclosure. The octagonal prism shape is maintained by a structure made of \SI{21}{\milli\meter} thick plywood. The plywood panels are assembled inside an extruded aluminum frame using aluminum fixtures and titanium and stainless steel bolts and nuts. The \SI{1}{\milli\meter} thick annealed permalloy sheets are stacked and pressed against the supporting plywood structure by means of titanium wood screws. High pressure laminate strips help distributing the load applied by the screws. Clearance holes for the screws in the permalloy sheets are laser cut after the annealing procedure. The gaps on the edges of the octagonal prism shape are closed using angled permalloy strips. The strips are bent after the annealing procedure. Angled and planar sheets are alternated as shown on figure~\ref{fig:stackup-example}. In the longitudinal direction, the permalloy sheets are close to \SI{3}{\meter} long. The few \si{\milli\meter} gap between consecutive planar sheets is bridged by offsetting the next sheet on the stack with \SI{50}{\%} overlap like in the inset of figure~\ref{fig:stackup-example}. In order to avoid saturation effects and help the magnetic equilibration field to penetrate effectively, the nominal thickness of the shells varies from \SI{3}{\milli\meter} at the ends up to \SI{8}{\milli\meter} in the central region. On the example of figure~\ref{fig:stackup-example}, the transition between three planar sheets (nominal thickness \SI{3}{\milli\meter}) and five planar sheets (nominal thickness \SI{5}{\milli\meter}) is depicted. The four end-caps are built following the same principles as the main section of the shield. The connection between the end-caps and the main body is done by overlapping the extremities of the permalloy sheets from both parts over \SI{20}{\centi\meter} and pressing them together against the plywood structure.

This construction allows precise positioning of each sheet while keeping the material stress-free and hence reducing build-up of inhomogeneities in its magnetic permeability. It is in particular crucial that the mechanical robustness is given by the plywood structure and not by the shielding material itself. Also, the plywood structure and the air gap between the layers provide an electrical isolation over \SI{10}{\mega\ohm}. This design is also fully scalable in length. The permalloy sheet material is annealed before assembly on the plywood structure, making the largest pieces to anneal only \SI{3}{\meter} long. This removes the need for a large hydrogen furnace\cite{Dickerson2012} irrespective of the final length of the shield. Finally, the possibility to open individual faces of the shells provides great access and flexibility for the installation of the vacuum chamber.

In order to minimize the free energy of the magnetic domains in the shielding material, we use a magnetic equilibration (``degaussing'') procedure similar to the one described by Altarev \textit{et al.}\cite{Altarev2014} We apply the equilibration field using coils routed around the faces of the octagonal prism shells using PVC pipes as guides (see figure~\ref{fig:open-shield}). For the inner shell, we use two sets of five turns of \SI{6}{\milli\meter\squared} copper wire connected in parallel, building up a coil enclosing each face of the permalloy octagonal prism twice (red and green lines on figure~\ref{fig:open-shield}). The resistive impedance of the equilibration coils for the inner layer is around \SI{2.5}{\ohm} per coil. We connect these coils in parallel to relax the voltage requirements on the coil driver. For the outer layer, we use only a single pass per face with five turns of the same wire which amounts to \SI{2.7}{\ohm} of resistive load. Finally, we set up another pair of coils for the end-caps with five turns of \SI{2.5}{\milli\meter\squared} cross-section wire per end-cap. These coils are wired in series and sum up to a resistance of \SI{0.3}{\ohm}. We drive the equilibration coils sequentially, first for the inner shell, then the end-caps, the outer shell, and finally the inner shell again. Each step starts by feeding a \SI{6}{\hertz}, \SI{10}{\ampere} RMS sinusoidal current in the coil, then decreasing the peak current linearly over \SI{100}{\second}. The full equilibration sequence therefore lasts around \SI{7}{\minute}. The target current waveform is calculated on a computer and fed into a \num{16}-bit voltage output DAC that drives an offset-trimmed current-mode amplifier. Residual offsets at the output of the current amplifier are further reduced by an external low distortion transformer.

\section{Measurements}

\begin{figure}
  \includegraphics[width=.48\textwidth] {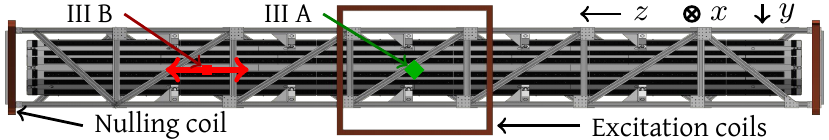}
  \caption{Setup for the characterization measurements. In section~\ref{sec:shielding-factor}, we use three orthogonal coil pairs around a fixed probe in the middle of the shield (\greendiamond) to measure the dynamical shielding factors in all three spatial directions. For the residual field measurements in section~\ref{sec:residual-field}, the probe (\redrectangle) travels along the longitudinal axis ($z$ direction) of the shield and we coarsely null the field at the ends of the shield using two constant current coils while the excitation coils are not used.}
  \label{fig:meas-setup}
\end{figure}

We determine the dynamical shielding factors in the middle of the shield (section~\ref{sec:shielding-factor}) and the residual vector field along the full longitudinal axis (section~\ref{sec:residual-field}) using the setup shown in figure~\ref{fig:meas-setup}. The end-caps are fully closed on both sides. We effectively reproduce the conditions of the final experimental apparatus by using a replicate of the vacuum chamber's aluminum pipe as a guide for the magnetic field sensors. For practicality reasons however, the shield is in horizontal position whereas it will be implemented vertically in the final configuration.

\subsection{Dynamical shielding factor}
\label{sec:shielding-factor}

\begin{figure}
  \includegraphics[width=.45\textwidth] {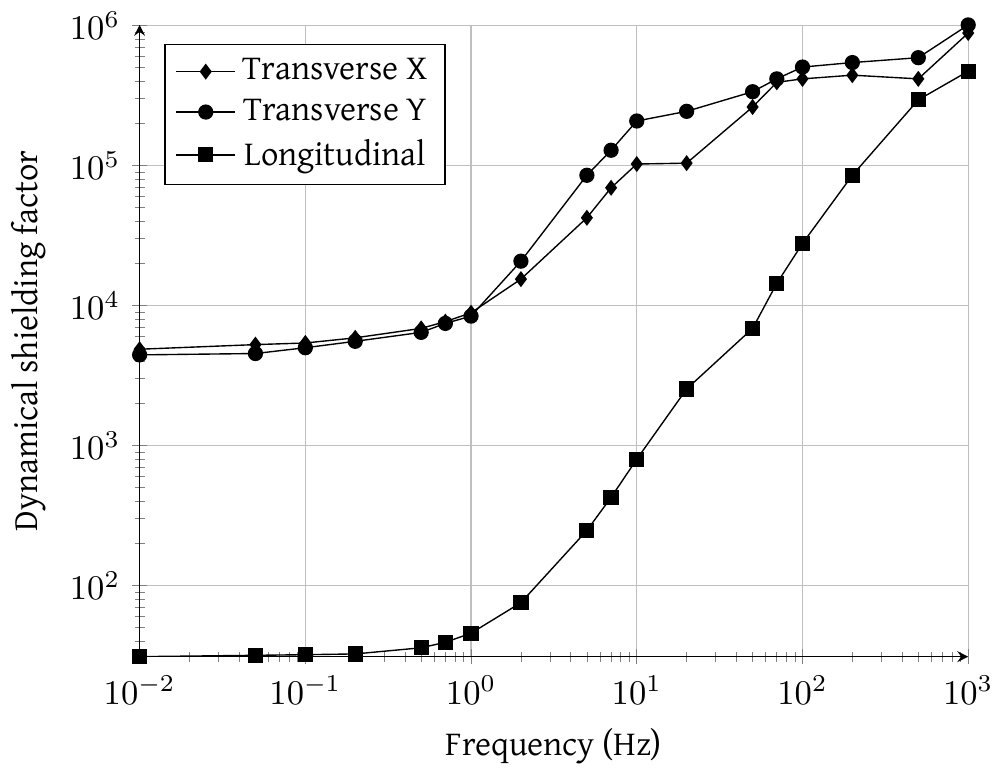}
  \caption{Measurement of the dynamical shielding factor at the center of the shield with end-caps installed. Effective eddy current shielding the aluminum pipe used as a guide for the probe is visible through the increased shielding factor at frequencies above \SI{1}{\hertz}.}
  \label{fig:shielding-factor}
\end{figure}

We first measure the response of the shield to a sinusoidal external field perturbation. We define the dynamical shielding factor as the ratio of the amplitude of the applied perturbation to the measured amplitude inside the shield.\cite{Altarev2014} We apply the external perturbation using a set of three calibrated rectangular coil pairs of dimensions $\SI{3}{\meter}\times\SI{2}{\meter}$ around the center of the shield. The resulting field in the corresponding frequency band is recorded by a three axes fluxgate sensor placed in the middle of the shield. To avoid saturating the magnetometer's digitizer, we vary the applied field between \SI{700}{\nano\tesla} and \SI{2000}{\nano\tesla}. Figure~\ref{fig:shielding-factor} shows the measured dynamical shielding factor versus frequency for all three directions. In the transverse directions, the shielding ratio reaches values above \num{4000} at \SI{0.01}{\hertz}, similar to other two-layer designs.\cite{Altarev2014} The corresponding longitudinal damping is more than \num{100} times lower, as expected from the large length-to-diameter ratio.\cite{Dickerson2012} Owing to the cancellation of the field's divergence in vacuum, the longitudinal field must be homogeneous if the transverse gradients are nulled and it can therefore be adjusted by a simple solenoid. Finally, we note that the change of slope observed in all directions around \SI{1}{\hertz} is characteristic of the crossover between effective shielding by the permalloy sheets at low frequencies (magnetostatic shielding) and by the aluminum pipe above this threshold (eddy current shielding\cite{Zimmerman1977,Stroink1981}).

\subsection{Residual field}
\label{sec:residual-field}

\begin{figure*}
  \includegraphics[width=.98\textwidth] {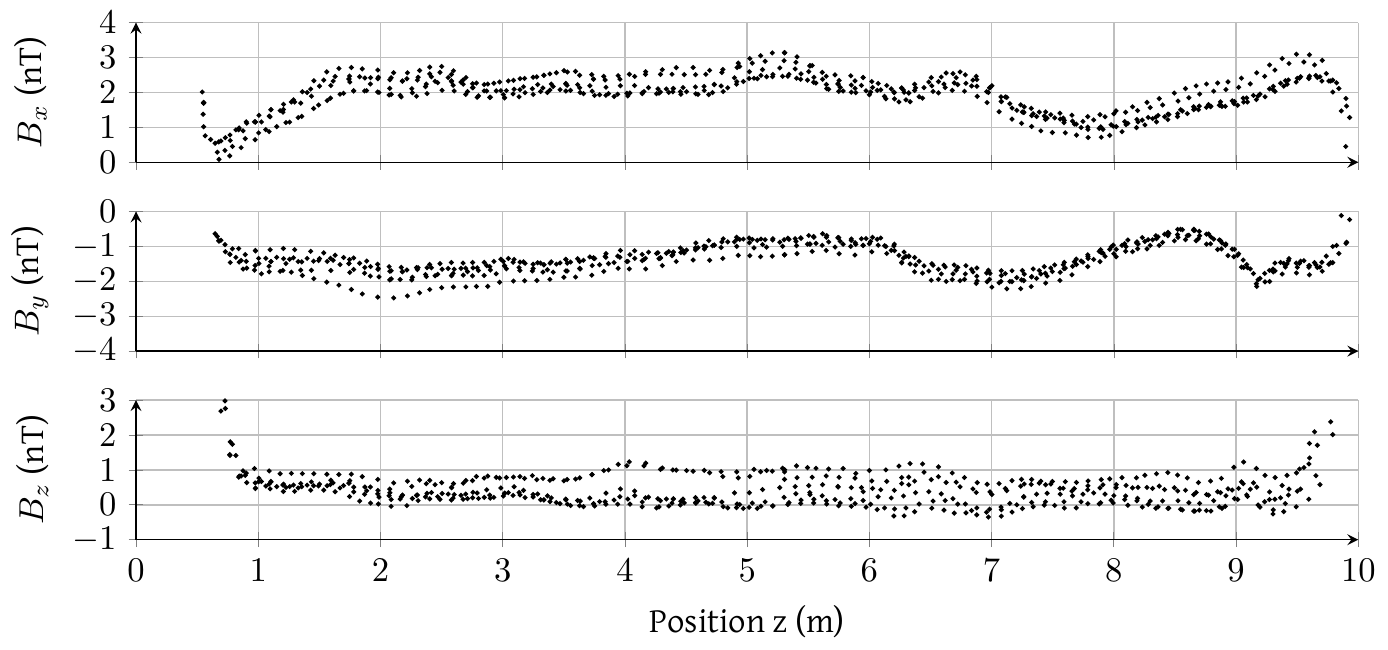}
  \caption{Measurements of the residual magnetic field components along the shield's symmetry axis. The magnitude of the residual field does not exceed \SI{4}{\nano\tesla} over the inner \SI{8}{\meter} with a reproducibility better than \SI{1.5}{\nano\tesla} between scans. The directions $x$ and $y$ are transverse while $z$ is longitudinal. The data consists of \num{570} points from \num{6} full length scans distributed over two consecutive days with magnetic equilibration between the scans.}
  \label{fig:field-map}
\end{figure*}

We also map the three components of the residual field along the symmetry axis of the shield. We mount a three axes magnetometer\footnote{Bartington Instruments, Model MAG-03, \SI{100}{\micro\tesla} full range} on a wooden mount assembled with non-metallic connectors to enable it travelling on the shield's axis guided by the aluminum pipe. We measure the position of the sensor in the shield using a remote controlled laser distance meter. We place two constant current loops around the end-caps of the magnetic shield to coarsely null the longitudinal field component near the entrance of the shield (figure~\ref{fig:meas-setup}). The shield is otherwise fully passive and we in particular did not implement any active external field stabilization. The magnetometer's output signals are conditioned by commercial pre-amplifiers\footnote{Bartington Instruments, Model PSU-1} and digitized using auto-zeroed \num{6.5} digits digital multimeters\footnote{Keysight Technologies, Model 34461A}. We perform six scans of the field on the shield's axis over two consecutive days. Each scan consists of ca. \num{90} points separated by ca. \SI{10}{\centi\meter} to map the full \SI{10}{\meter} of shielded region. Each point is a \SI{20}{\second} average, corresponding to \num{50} acquisition cycles from the digital multimeters\footnote{\num{50} points with \num{10} powerline cycles long acquisition windows and auto-zero feature turned on}. Since the probe offsets are only calibrated to \SI{+-5}{\nano\tesla} in the factory, we perform a custom offset calibration step between all scans. For this, we measure both the parallel and antiparallel components of the field and take their average to find the probe offset. This is better realized in low field regions to limit errors due to imperfect inversion of the sensitive axis direction. We observe that measured offsets are reproducible at the one nanotesla level and therefore attribute an uncertainty of \SI{+-500}{\pico\tesla} to the absolute magnetic field measurements. We note however that probe offsets can vary more significantly when cycling the sensor's power or stressing readout connectors.\footnote{With a \SI{100}{\micro\tesla} full scale sensor, \SI{1}{\nano\tesla} corresponds to \SI{100}{\micro\volt} which can easily appear spuriously on a faulty electrical connection.} Finally, to account for the large variability of the external field, up to \SI{400}{\nano\tesla} peak-to-peak on the \SI{1}{\hour} scale, due to industrial activity in the surrounding area, we also apply the equilibration procedure between each run. Figure~\ref{fig:field-map} shows the measured field maps with a residual magnetic field below \SI{4}{\nano\tesla} over the inner \SI{8}{\meter} of the shielded region. The reproducibility of the scans over \SI{20}{\centi\meter} windows is better than \SI{1.5}{\nano\tesla} max-to-min and better than \SI{500}{\pico\tesla} on the standard deviation, demonstrating the robustness of the magnetic equilibration procedure.

\section{Application to precision atomic accelerometry}

For our application in precision atom interferometry, the leading systematic error associated with magnetic fields is due to the Zeeman effect. For atomic states with zero magnetic quantum number, the linear component of the Zeeman effect vanishes and the potential seen by the atoms due to a longitudinal magnetic field profile $B(z)$ is quadratic in the total field strength:

\begin{equation}
  \label{eq:zeeman-potential}
  V(z) = -\frac{1}{2}\hbar\alpha B(z)^2.
\end{equation}

Here, $\hbar$ is the reduced Planck constant and $\alpha$ the atomic species' clock transition Zeeman coefficient (\SI{57.5}{\giga\hertz\per\tesla\squared} for the \Rb{87} atom\cite{Steck2015}). Using perturbation theory,\cite{Chiu1980,Ufrecht2019} we evaluate the phase $\phi$ of an atom interferometer in the presence of an arbitrary but small perturbation potential $V(z)$. To first order, we get

\begin{equation}
  \label{eq:ifo-phase}
  \phi = \phi_0 - \frac{1}{\hbar}\oint\mathrm{d}t\,V(z_0(t))
\end{equation}

where $\phi_0$ is the phase of the corresponding unperturbed interferometer and the integral spans over the oriented loop formed by the unperturbed classical trajectories $z_0(t)$. In the presence of a local gradient with no curvature and writing the magnetic field profile as $B(z) = B_0 + \epsilon b(z)$, $\epsilon\ll 1$, the local bias acceleration on atoms of mass $m$ reads

\begin{equation}
  \label{eq:simple-bias-acceleration}
  \delta a = \epsilon\frac{\hbar\alpha B_0}{m}\frac{\partial b(z)}{\partial z} + \mathcal{O}(\epsilon^2).
\end{equation}

We numerically calculate the magnetic field gradient from the data of figure~\ref{fig:field-map}. The resolution is limited by the finite spatial sampling of the data. However, between all six scans presented in figure~\ref{fig:field-map}, the sensor positions were not exactly reproduced, making the sampling grid effectively finer than \SI{10}{\centi\meter}. Moreover, due to smoothness requirements of the magnetic field, we do not expect spurious features to have been missed in the residual field measurement and therefore interpolate the data linearly and take the local slope as the local gradient. Over the inner \SI{8}{\meter} of shielded region, the local longitudinal magnetic field gradient never exceeds \SI{3}{\nano\tesla\per\meter} which corresponds to a maximum local acceleration of \SI{1.2}{\pico\meter\per\second\squared} for \Rb{87} atoms when $B_0 = \SI{1.5}{\micro\tesla}$. This effect can be constrained further by applying the perturbation theory result of equation~\ref{eq:ifo-phase} since the integral over the unperturbed classical trajectories smoothes local spikes in the magnetic field profile. For a simple drop mode operation, we find the bias for an interferometer spanning the inner \SI{8}{\meter} of shielded region to be smaller than five parts in $10^{15}$ of the Earth's local gravitational acceleration.

\section{Conclusion}

We reported on the design, construction, and characterization of a \SI{10}{\meter} high-performance magnetic shield with application in Very Long Baseline Atom Interferometry, achieving residual fields below \SI{4}{\nano\tesla} and longitudinal gradients smaller than \SI{2.5}{\nano\tesla\per\meter} over the central \SI{8}{\meter}. Owing to the use of pre-annealed permalloy sheet material in an octagonal prism geometry and careful, stress-free assembly, the design is fully scalable in length, effectively removing the need for a large scale hydrogen furnace for annealing while improving the homogeneity of the shielded region's magnetic field by an order of magnitude compared to previous work.\cite{Dickerson2012} This namely opens shielding possibilities for ultra large scale experiments proposed to detect gravitational waves or search for exotic matter with atomic matter-waves \cite{Zhan2019,Coleman2018} where monolithic designs cannot be considered. For interferometer geometries using the full length of the baseline, our shield leads to a Zeeman effect associated bias for \Rb{87} atoms below five parts in $10^{15}$ of the Earth's local gravitational acceleration. This enables a new class of absolute gravimeters for long-term gravity monitoring and reference networks.\cite{VanCamp2017} Finally, when comparing the acceleration of two different atomic species, this paves the way towards Galilean tests of the universality of free fall with atomic test masses beyond the $10^{-13}$ level.\cite{Hartwig2015}

\section*{Acknowledgements}

The Hannover Very Long Baseline Atom Interferometry facility is a major research equipment funded by the German Research Foundation (Deutsche Forschungsgemeinschaft, DFG). This work is supported by the Collaborative Research Centers 1128 ``geo-Q'' (project A02) and 1227 ``DQ-mat'' (project B07), and Germany's Excellence Strategy within EXC-2123 ``QuantumFrontiers'' (project number 390837967). D. S. acknowledges funding from the German Federal Ministry of Education and Research (BMBF) through the funding program Photonics Research Germany (contract number 13N14875). E.~W. acknowledges support from ``Nieders\"achsisches Vorab'' through the ``Quantum- and Nano-Metrology~(QUANOMET)'' initiative (project QT3). St.~S. and P.~F. acknowledge support by EXC-153 ``Origin and Structure of the Universe'' (project number 24799710). The work of C.~U. is supported by the German Aerospace Center (DLR) with funds provided by the Federal Ministry for Economic Affairs and Energy (BMWi) due to an enactment of the German Bundestag under grants no. DLR~50WM1556 and 50WM1956. E.~W., D.~T., E.~M.~R., and D.~S. thank I. Fan and A. Schnabel for valuable insight at the beginning of the project. E.~W. thanks T.~Wendrich, K.~M.~Knaak, T.~Hensel, and T.~Rehmert for logistics help and the loan of measurement equipment.

\bibliography{paper}

\end{document}